\documentclass[useAMS,usenatbib]{mn2e}
 \usepackage{graphics}
 \usepackage{subfigure}
 \usepackage{natbib}
 \usepackage{upgreek}
\usepackage{bibentry}
 \usepackage{url}
 \usepackage{epsfig}
\usepackage{epstopdf}

\usepackage{amssymb}
\usepackage{amsmath}
\usepackage{aas_macros}

\title[Orbital Dynamics of Exoplanetary Systems Kepler-62, HD 200964 and  Kepler-11]{Orbital Dynamics of 
Exoplanetary Systems Kepler-62, HD 200964 and  Kepler-11}
\author[Rajib Mia and Badam Singh Kushvah]{Rajib Mia\thanks{E-mail:
rajibmia.90@gmail.com (RM); bskush@gmail.com (BSK).} and Badam Singh Kushvah$^{\star}$\\
Department of Applied Mathematics, Indian School of Mines, Dhanbad 826004, Jharkhand, India 
}

\begin{document}

\date{}
\volume{457}
\pagerange{1089--1100} \pubyear{2016}
\maketitle

\label{firstpage}
\begin{abstract}
The presence of mean-motion resonances (MMRs) in exoplanetary systems is a new
exciting field of celestial mechanics which motivates us to consider this work to  study   the dynamical behaviour of  exoplanetary systems by time evolution of the orbital elements of the planets. Mainly, we study the
influence of planetary perturbations on semimajor axis and eccentricity. 
We identify $(r+1):r$ MMR terms in the expression of disturbing function and obtain  the perturbations from the truncated disturbing function.
Using the expansion of the disturbing function of three-body problem and an analytical approach, we solve the equations of motion. The solution which is obtained analytically is compared with that of obtained by numerical method to validate our analytical result.
In this work, we consider three exoplanetary systems namely Kepler-62, HD 200964 and Kepler-11. We have plotted the evolution of the resonant angles and found that they librate around constant value. In view of this, our opinion is that two planets of each system Kepler-62, HD 200964 and Kepler-11 are in 2:1, 4:3 and 5:4  mean motion resonances, respectively.
 \end{abstract} 
\section{Introduction} 
The exoplanets are  planets outside our Solar system and which  have been
detected since $1989$. The first multiple exoplanet system was discovered by 
\cite{Wolszczan1992Natur.355..145W} and found two very low-mass objects orbiting
the pulsar PSR B1257+12 with the help of pulse timing methods.
Over the last decades, the discovery of exoplanetary systems  increases. 
More than $1000$ exoplanets (see e.g., http:/www.exoplanet.eu) have been discovered by 2015 December. 
Among these known exoplanets, there is a total of 504 known multiplanetary systems where each star has at least two confirmed planets and some system with more than one planet have near resonant orbital configurations
(see e.g., http://exoplanet.eu/catalog/, an online data base for exoplanetary systems developed by \cite{Schneider2011A&A...532A..79S}).
In our Solar system  and similar exoplanetary systems, mean motion resonances(MMRs) are common feature \citep{Mustill2011IAUS..276..300M}. MMRs occur when two orbital periods are near a ratio of two small integers \citep{Petrovich2013ApJ...770...24P}, and the resonant argument(certain combination of orbital angles) is librating.
In our own Solar system 5:2, 3:1 and 2:1 near MMRs occur in
Jupiter--Saturn, Saturn--Uranus and Uranus--Neptune systems. In the exoplanetary
systems, only a handful of multiplanetary systems contain a pair of
planet in MMRs. The majority of which are  in 2:1 MMR
\citep{beauge2006planetary}. For example, GL 876, HD 82943 and HD 128311 are in
2:1 MMR. Also others found in 3:2 \citep{Malhotra1993ApJ...407..266M}, 4:3 MMR
\citep{santos2014formation} etc. 

A lot of work has performed regarding the stability and instability of the
Lagrangian points using circular or elliptical restricted three-body problem  \citep{Szebehely1967torp.book.....S,Papadakis2005Ap&SS.299...67P,
Pal2015MNRAS.446..959P}, depending on the mass ratio of the primaries and the
eccentricities of the orbits.

In astronomy, the habitable zone (HZ) is a region around a star in which water can exist permanently in the liquid state at the 
 surface of the planet. Although there are more than one thousand exoplanets, among them only a dozen of planets have been confirmed in the HZ. Kepler-62e is a super-Earth like exoplanet \citep{borucki2013kepler} discovered in orbit around star Kepler-62. This exoplanet was found using the transit method. Kepler-62e is most likely a terrestrial planet in the inner part of its host star's HZ and it has Earth similarity index 0.83. It is roughly 60 percent larger than Earth. This recently discovered system has five planets. Among these five planets, two of them Kepler-62e and Kepler-62f are supposed to be in the 2:1 MMR \citep{borucki2013kepler}.
The star HD 200964 have two giant exoplanets and they are separated by only 0.35 AU. The planets are in a 4:3 MMR \citep{Johnson2011AJ....141...16J}.
Kepler-11 is a Sun-like star, located approximately 2000 light years far from Earth. It is the first exoplanetary system consisting six transiting planets. Although, none of these planets be in low-ratio orbital resonances, Kepler-11b and Kepler-11c are in 5:4 MMR \citep{lissauer2011closely}.

\cite{Malhotra1993ApJ...407..266M} presented a detail theoretical analysis of
three-body effects in the putative planetary system of PSR 1257+12. She provided
explicit elements for the time dependence of the osculating elements that are
needed for an improved timing model for the system. She  has also shown that
the 3:2 MMR of the orbital periods affect periodic variations
of the Keplerian orbital parameters. She has obtained the expansion of
disturbing function and identified only two first-order 3:2 resonance term. But we
identify $(r+1):r$ MMR terms in the expression of disturbing
function and obtained  the perturbations from the truncated disturbing function.
In this manuscript, we have taken three exoplanetary systems namely
Kepler-62, HD 200964 and Kepler-11 and as per our knowledge they contain planets
in 2:1, 4:3 and 5:4 near MMRs, respectively. We apply theory
discussed in Sections \ref{sec:2sec2} and \ref{sec:2sec3} individually in our
three exoplanetary systems that we have chosen.

This paper is organized as follows. In Section 2, we introduce the general three-body problem and the
stability of exoplanetary system. The expansions of the disturbing function and perturbation equations of orbital elements are introduced in Section 3. In Section 4, we present  our secular resonance dynamics of exoplanetary system. Applications of the model for the case of 2:1 resonance are shown in Section 5 for Kepler-62 system. The case of the 4:3 resonance is discussed in Section 6 for HD 200964 system. In Section 7, the case of 5:4 is discussed for Kepler-11 system. Finally, Section 8 is devoted to conclusions.
\section{ General three body-problem and dynamical Stability}\label{sec:2}
Suppose the barycentric position vectors of star and two planets be $\vec{R_{\star}}$, $\vec R_1$ and $\vec R_2$ respectively. Let two planets of masses $M_1$ and $M_2$ orbiting a star of mass $M_\star$. Let  the Jacobi coordinates $\vec r_1$ and $\vec r_2$ be the position of $M_1$ relative to $M_\star$ and position of $M_2$ relative to the centre of mass $M_\star$ and $M_1$.  We notice that this system is different from the restricted problems in which one of the bodies must have negligible mass. Also we ignore the effects of oblateness of star and the planets. We assume that the planets in all system are coplanar. This approximation of coplanarity makes sense because for exoplanets, their mutual inclination is not known with any precision, and is taken equal to zero \citep{beauge2003extrasolar}. Let $a_i, e_i, \lambda_i$ and $\omega_i$ be the semimajor axis, eccentricity, mean longitude and  the longitude of the pericentre of the $i$th planet for $i=1,2.$ 

In the three-body problem,  one of the interesting topic is the stability of the Solar system or exoplanetary system. Although, in celestial mechanics it is the unsolved problem till now, many scientist studied this because of its importance in many areas like space science, astronomy and astrophysics. In the context of planets and exoplanets formulation, \cite{Graziani1981ApJ...251..337G} studied conditions under which model planetary systems which consisting of a star and two planets with coplanar and initially circular orbits. Based on their results, they obtained a necessary condition for orbital stability  
\begin{eqnarray}
\mu=0.5\frac{M_1+M_2}{M_{\star}}<\mu_{\text{crit}}=0.175\frac{\Delta^3}{(2-\Delta)^{\frac{3}{2}}}, \quad \mu\leq 1,
\end{eqnarray}
where $M_1$ and $M_2$ are masses of planetary companions and $M_{\star}$ is the mass of the star. The parameter $\Delta$ is the ratio of the separation distance between the companions of their mean distance from the star. Specifically they defined 
\begin{eqnarray}
\Delta=2\frac{R-1}{R+1}, \qquad R=\frac{a_2}{a_1},
\end{eqnarray}
where $a_1$ and $a_2$ are the semimajor axes of the inner and outer orbits, respectively.
The system will be unstable for the case $\mu\geq \mu_{\text{crit}}$ within a few tens of planetary orbits. 

In recent years, several exoplanetary systems have been discovered and it is very interesting to know their dynamical stability. There are several authors \citep{Fabrycky2010ApJ...710.1408F,Couetdic2010A&A...519A..10C,Davies2014prpl.conf..787D,adams2015stability,Petrovich2015ApJ...808..120P} who have studied about the stability of exoplanetary systems. 
\cite{Davies2014prpl.conf..787D} have reviewed the long-term dynamical evolution of the planetary systems. They have discussed the planet--planet interactions that take place within our own Solar system and in more tightly--packed planetary systems. Some system becomes dynamically unstable because of planet--planet interactions build up and this lead to strong encounters and ultimately either ejections or collisions of planets. It was shown that the Solar system is chaotic, but the four giant-planet sub-system of the Solar system is stable although the terrestrial-planet sub-system is marginally unstable with a small change of planet--planet encounters during the lifetime of the Sun. 

Likewise here we discuss the dynamical stability of two-planet sub-system of Kepler-62 system, HD 200964 system and two-planet sub-system of Kepler-11 system.
\cite{Petrovich2015ApJ...808..120P} provides an independent review on the stability of two-planet systems. They studied the dynamical stability and fates of hierarchical (in semimajor axis) two-planets systems with arbitrary eccentricities and mutual inclinations. They proposed the following new criteria for dynamical stability
\begin{eqnarray}
&&r_{\text{ap}}=\frac{a_{\text{out}}(1-e_{\text{out}})}{a_{\text{in}}(1+e_{\text{in}})}>Y,\label{eq:3eqn}
\end{eqnarray}
where,
\begin{eqnarray}
&&Y=2.4[\text{max}(\mu_{\text{in}},\mu_{\text{out}})]^{\frac{1}{3}}(\frac{a_{\text{out}}}{a_{\text{in}}})^{\frac{1}{2}}+1.15,\nonumber
\end{eqnarray}
$a_{\text{in}}$ and $a_{\text{out}}$ are the semimajor axes of the inner and outer planet and $\mu_{\text{in}}$ and $\mu_{\text{out}}$ are the planet-to-star mass ratios of the inner and outer planets, respectively.
The hierarchical two planets systems are stable if they satisfy the condition in equation (\ref{eq:3eqn}) and systems that do not satisfy equation (\ref{eq:3eqn}) are expected to be unstable. The fate of the unstable systems classified according to planetary masses 
as when $\mu_{\text{in}}>\mu_{\text{out}}$, system lead to ejections and for $\mu_{\text{in}}<\mu_{\text{out}}$, there is a slightly favouring of collisions with the star. 

Now the goal is to apply this latest stability criteria of \cite{Petrovich2015ApJ...808..120P} to Kepler-62, HD 200964, and Kepler-11 system.
For the case of Kepler-62 system considering only two planets Kepler-62e and Kepler-62f, using the masses and initial orbital elements, we obtain $r_{\text{ap}}=1.34767$, $Y=1.31762$ with $\mu_{\text{in}}>\mu_{\text{out}}$. Hence, stability criteria in equation (\ref{eq:3eqn}) implies that system is stable.  
In the planetary system HD 200964, the inner and outer planets HD 200964b and HD 200964c are in eccentric orbit with $e_{\text{in}}\simeq0.040$, $e_{\text{out}}\simeq0.181$ and $a_{\text{in}}\simeq1.601$ AU, $a_{\text{out}}\simeq1.950$ AU. 
Using the initial orbital elements and masses we found $r_{\text{ap}}=0.959166$ and $Y=1.43346$ with $\mu_{\text{in}}>\mu_{\text{out}}$. So, stability criteria in equation (\ref{eq:3eqn}) implies that HD 200964 system is unstable against ejections. Also for the exoplanetary system Kepler-11 considering only two planets, Kepler-11b and Kepler-11c, we found
$r_{\text{ap}}=1.059964>Y=1.24015$ with $\mu_{\text{in}}<\mu_{\text{out}}$. Hence, this system is unstable and there is a slightly favouring of collisions with star than ejections. 

 \section{Expansion of the disturbing function and perturbation equations of Orbital elements}\label{sec:2sec2}
The expansion of the disturbing function in the orbital eccentricities to first order for the periodic 
terms and to second order for the secular terms is given as \citep{Malhotra1993ASPC...36...89M} 
 \begin{eqnarray} \nonumber
 a_2V&=&Q(\psi,\alpha)-\alpha\cos\psi-e_1\cos(\lambda_1-\omega_1)\left[\alpha\frac{\partial}{\partial\alpha}Q(\psi,\alpha)
 \right.\\&&\left. \nonumber
 +\alpha\cos\psi \right]+ e_1\sin(\lambda_1-\omega_1)\left[2\frac{\partial}{\partial\psi}Q(\psi,\alpha)+2\alpha\sin\psi\right]\nonumber\\&& \label{eq:1}
 +e_2\cos(\lambda_2-\omega_2)\left[(1+\alpha\frac{\partial}{\partial\alpha})Q(\psi,\alpha)-2\alpha\cos\psi\right]\\&&
 -e_2\sin(\lambda_2-\omega_2)\left[2\frac{\partial}{\partial\psi}Q(\psi,\alpha)+2\alpha\sin\psi\right]\nonumber\\&&
 +\frac{1}{8}\alpha\left[b_{\frac{3}{2}}^{(1)}(\alpha)(e_1^2+e_2^2)-2b_{\frac{3}{2}}^{(2)}(\alpha)e_1e_2\cos(\omega_1-\omega_2)\right],\nonumber
 \end{eqnarray}
 where
 \begin{eqnarray}
 \lambda_j&=&\mathcal{M}_j+\omega_j,\nonumber\\
 \psi&=&\lambda_1-\lambda_2,\\\label{eq:2}
 Q(\psi,\alpha)&=&(1-2\alpha\cos\psi+\alpha^2)^{-\frac{1}{2}}.\nonumber
 \end{eqnarray}
 Also the Laplace coefficients $b_s^{(j)}(\alpha)$ and Fourier series expansion of $Q(\psi,\alpha)$ are defined as
 \begin{eqnarray}
 &&\frac{1}{2}b_s^{(j)}(\alpha)=\frac{1}{2\uppi}\int _0^{2\uppi}\frac{\cos jp \ dp}{(1-2\alpha\cos p+\alpha^2)^s},\\&&\label{eq:3}
  Q(\psi,\alpha)=\frac{1}{2}\sum_{j=-\infty}^{\infty} b_s^{(j)}(\alpha)\cos j\psi.\label{eq:4}
 \end{eqnarray}
From equations \eqref{eq:1} and \eqref{eq:4}, we can determine two terms associated with the two first-order $r+1:r$  arguments, namely $\theta_1=(r+1)\lambda_2-r\lambda_1-\omega_1$ and 
 $\theta_2=(r+1)\lambda_2-r\lambda_1-\omega_2$, where $\lambda_1$ and $\lambda_2$ are the mean 
 longitudes of planet 1 and planet 2, respectively. Therefore, the truncated disturbing function is
\begin{eqnarray}
a_2V&=&Q(\psi,\alpha)-\alpha\cos\psi\nonumber\\&&
+K_1(\alpha)e_1\cos\{(r+1)\lambda_2-r\lambda_1-\omega_1\}\nonumber\\&&
+K_2(\alpha)e_2\cos\{(r+1)\lambda_2-r\lambda_1-\omega_2\}\nonumber\\&&
+\frac{1}{8}\alpha \left\{b_{\frac{3}{2}}^{(1)}(\alpha)(e_1^2+e_2^2)
-2b_{\frac{3}{2}}^{(2)}(\alpha)e_1e_2
\nonumber\right.\\&&\left.
\times\cos(\omega_1-\omega_2)\right\},\label{eq:d1}
\end{eqnarray}
where 
\begin{eqnarray}
K_1(\alpha)&=&-\left\{(r+1)+\frac{\alpha}{2}D\right\}b_{\frac{1}{2}}^{(r+1)}(\alpha),\\
K_2(\alpha)&=&\left\{(r+\frac{1}{2})+\frac{\alpha}{2}D\right\}b_{\frac{1}{2}}^{(r)}(\alpha)
,\qquad D=\frac{d}{d\alpha}.
\end{eqnarray}
Now using the truncated disturbing function of Eq.\eqref{eq:d1}, we obtain the perturbation equations
for the time variation of the orbital elements as:
the time variation of the semi-major axes are 
\begin{eqnarray}
\frac{\dot{a_1}}{a_1}&=&2\frac{M_2}{M_\star}n_1\alpha\left[\partial_{\psi}Q(\psi,\alpha)+\alpha\sin\psi
\nonumber\right.\\&&\left. \label{eq:2pa1}
-r\left(r+1+\frac{\alpha}{2}D\right)b_\frac{1}{2}^{(r+1)}(\alpha)e_1
\nonumber\right.\\&&\left.
\times\sin\{(r+1)\lambda_2-r\lambda_1-\omega_1\}
\nonumber\right.\\&&\left. 
+r\left(r+\frac{1}{2}+\frac{\alpha}{2}D\right)b_{\frac{1}{2}}^{(r)}(\alpha)e_2
\nonumber\right.\\&&\left.
\times\sin\{(r+1)\lambda_2-r\lambda_1-\omega_2\}\right],
\end{eqnarray}
\begin{eqnarray}
\frac{\dot{a_2}}{a_2}&=&-2\frac{M_1}{M_\star}n_2\left[\partial_{\psi}Q(\psi,\alpha)+\alpha\sin\psi
\nonumber\right.\\&&\left. \label{eq:2pa2}
-(r+1)\left(r+1+\frac{\alpha}{2}D\right)b_\frac{1}{2}^{(r+1)}(\alpha)e_1
\nonumber\right.\\&&\left.
\times\sin\{(r+1)\lambda_2-r\lambda_1-\omega_1\}
\nonumber\right.\\&&\left. 
+(r+1)\left(r+\frac{1}{2}+\frac{\alpha}{2}D\right)b_{\frac{1}{2}}^{(r)}(\alpha)e_2
\nonumber\right.\\&&\left.
\times\sin\{(r+1)\lambda_2-r\lambda_1-\omega_2\}\right].
\end{eqnarray}
The time variation of the eccentricities:
\begin{eqnarray}\label{eq:2p1}
\dot{e_1}&=&\frac{M_2}{M_\star}n_1\alpha\left[\frac{1}{4}\alpha b_{\frac{3}{2}}^{(2)}(\alpha)e_2\sin(\omega_1-\omega_2)
\nonumber\right.\\&&\left. 
-\left(r+1+\frac{\alpha}{2}D\right)b_\frac{1}{2}^{(r+1)}(\alpha)
\nonumber\right.\\&&\left. 
\times\sin\{(r+1)\lambda_2-r\lambda_1-\omega_1\}
\right],\label{eq:2p6}
\end{eqnarray}
\begin{eqnarray}\label{eq:2p2}
\dot{e_2}&=&-\frac{M_1}{M_\star}n_2\left[\frac{1}{4}\alpha b_{\frac{3}{2}}^{(2)}(\alpha)e_1\sin(\omega_1-\omega_2)
\nonumber\right.\\&&\left. 
-\left(r+\frac{1}{2}+\frac{\alpha}{2}D\right)b_\frac{1}{2}^{(r)}(\alpha)
\nonumber\right.\\&&\left. 
\times\sin\{(r+1)\lambda_2-r\lambda_1-\omega_2\}
\right].\label{eq:2p7}
\end{eqnarray}
The time variation of the periastrons:
\begin{eqnarray}\label{eq:2p3}
\dot{\omega_1}&=&\frac{M_2}{M_\star}n_1\alpha\left[\frac{1}{4}\alpha b_{\frac{3}{2}}^{(1)}(\alpha)-\frac{1}{4}\alpha b_{\frac{3}{2}}^{(2)}(\alpha)\frac{e_2}{e_1}\cos(\omega_1-\omega_2)
\nonumber\right.\\&&\left.
-\frac{1}{e_1}\left(r+1+\frac{\alpha}{2}D\right)b_{\frac{1}{2}}^{(r+1)}(\alpha)
\nonumber\right.\\&&\left.
\times\cos\{(r+1)\lambda_2-\lambda_1-\omega_1\}\right],\label{eq:2p8}
\end{eqnarray}
\begin{eqnarray}\label{eq:2p4}
\dot{\omega_2}&=&\frac{M_1}{M_\star}n_2\left[\frac{1}{4}\alpha b_{\frac{3}{2}}^{(1)}(\alpha)-\frac{1}{4}\alpha b_{\frac{3}{2}}^{(2)}(\alpha)\frac{e_1}{e_2}\cos(\omega_1-\omega_2)\label{eq:2p9}
\nonumber\right.\\&&\left.
+\frac{1}{e_2}\left(r+\frac{1}{2}+\frac{\alpha}{2}D\right)b_{\frac{1}{2}}^{(r)}(\alpha)
\nonumber\right.\\&&\left.
\times\cos\{(r+1)\lambda_2-r\lambda_1-\omega_2\}\right].
\end{eqnarray}

If we put $r=2$ in the above results, disturbing function and the time variations of orbital elements agree with that of \cite{Malhotra1993ApJ...407..266M}.
\section{Secular resonance Dynamics of Exoplanetary system}\label{sec:2sec3}
In this section  we discuss the long-term variations of the eccentricities of the exoplanets by secular theory with MMR. It is convenient to define the vertical and horizontal components of eccentricity by
\begin{eqnarray}
&&p_j=e_j\sin\omega_j, \qquad  q_j=e_j\cos\omega_j.
\end{eqnarray} 
These new variables have the advantage that they can remove the singularities at zero eccentricity in 
Eqs.\eqref{eq:2p1}-\eqref{eq:2p9}. After some calculation, we obtain the equations for the variation of
$p_j$ and $q_j(j=1,2)$ as
\begin{eqnarray}
&&\dot{p_j}=\sum_{k=1}^{2}A_{jk}q_{k}+E_j\cos\{(r+1)\lambda_2-r\lambda_1\},\\
&&\dot{q_j}=-\sum_{k=1}^{2}A_{jk}p_{k}-E_j\sin\{(r+1)\lambda_2-r\lambda_1\}.
\end{eqnarray}
These are the first-order differential equations and hence the problem of secular perturbations reduces
to the eigenvalue problem, where the coefficient matrix $A$ is given by
\[ \left( \begin{array}{cc}
A_{11} & A_{12} \\
A_{21} & A_{22} \end{array} \right)\] 
and
\begin{eqnarray}
&&A_{11}=\frac{M_2}{4M_{\star}}n_1\alpha^2 b_{\frac{3}{2}}^{(1)}(\alpha), \quad  A_{12}=-\frac{M_2}{4M_{\star}}n_1\alpha^2 b_{\frac{3}{2}}^{(2)}(\alpha),\nonumber\\&&
A_{21}=-\frac{M_1}{4M_{\star}}n_2\alpha b_{\frac{3}{2}}^{(2)}(\alpha), \quad A_{22}=\frac{M_1}{4M_{\star}}n_2\alpha b_{\frac{3}{2}}^{(1)}(\alpha).
\end{eqnarray}
Also \begin{eqnarray}
&&E_1=-\frac{M_2}{M_\star}n_1\alpha\left((r+1)+\frac{\alpha}{2}D\right)b_{\frac{1}{2}}^{(r+1)}(\alpha),\nonumber\\ &&
E_2=\frac{M_1}{M_\star}n_2\left((r+\frac{1}{2})+\frac{\alpha}{2}D\right)b_{\frac{1}{2}}^{(r)}(\alpha).
\end{eqnarray}
The solutions are given by the superposition of free oscillations and forced oscillations as
\begin{eqnarray}
p_j(t)&=&\sum_{i=1}^2e_{ji}\sin(g_it+\beta_i )+F_j\sin((r+1)\lambda_2-r\lambda_1),\nonumber\\
q_j(t)&=&\sum_{i=1}^2e_{ji}\cos(g_it+\beta_i )+F_j\cos((r+1)\lambda_2-r\lambda_1),\label{eq:2p10}
\end{eqnarray} 
where the frequencies $g_i  \ (i=1,2)$ are the eigenvalues of the coefficient matrix $A$ and $e_{ji}$ 
are the component of the two corresponding eigenvectors. The normalization of eigenvectors and 
phases $\beta_i$ can be determined by the initial conditions. The amplitude of forcing is given as 
\begin{equation}\label{eq:f}
 \left( \begin{array}{cc}
F_1  \\
F_2\end{array} \right)=-B^{-1}.\left( \begin{array}{cc}
E_1  \\
E_2\end{array} \right),
\end{equation}
where $B=[A-\{(r+1)n_2-r n_1\}I]$ and I denotes a $2\times 2$ identity matrix.
Now if there is no MMR, then equation \eqref{eq:2p10} reduces to 
\begin{eqnarray}
&&p_j(t)=\sum_{i=1}^2e_{ji}\sin(g_it+\beta_i ),\nonumber\\&&
q_j(t)=\sum_{i=1}^2e_{ji}\cos(g_it+\beta_i ).\nonumber
\end{eqnarray} 
 which are similar to the classical Laplace--Lagrange secular solutions for the eccentricities \citep{Murray2000ssd..book.....M}.
 We present variations of the semimajor axes and eccentricities of Kepler-62, HD 200964 and Kepler-11 systems in following sections.
\begin{table}
 \begin{minipage}{120mm}
\caption{Orbital parameters of the  Kepler-62 system. \newline The data are taken from  
 \citet{borucki2013kepler}.\label{tab:2p1} } 
  \begin{tabular}{@{}rrrrrrrrrrr@{}}
  \hline
  Parameter &Kepler-62e & Kepler-62f&  \\\hline
 $P$(days)&$122.3874\pm 0.0008$&$267.291\pm0.005$\\           
 $a$(AU)&$0.427\pm0.004$&$0.718\pm0.007$ \\ 
 $i$(deg)&$89.98\pm0.02$&$89.90\pm0.03$\\
 $T_0$(BJD--2454900)&$83.404\pm0.003$&$522.710\pm0.006$\\
 $e\cos\omega$&$0.05\pm0.17$&$-0.05\pm0.14$\\
 $e\sin\omega$&$-0.12\pm0.02$&$-0.08\pm0.10$\\
 \hline
\end{tabular}
\end{minipage}
\end{table}
 \section{Kepler-62 system}\label{sec:2sec5}
 \subsection{The 2:1 MMR}
\begin{figure*}
\includegraphics[height=0.20\textwidth]{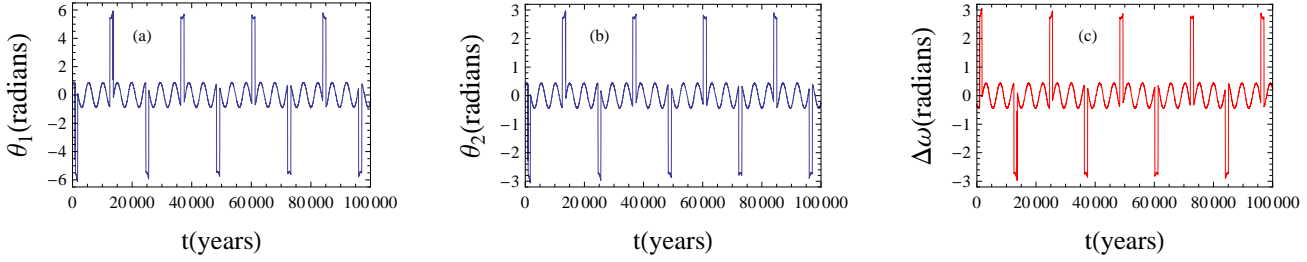}
\caption{The evolution of the resonant angles $\theta_1$ (a), $\theta_2$ (b) and apsidal angle $\Delta\omega$ (c) of Kepler-62 system. Note that $\theta_1$, $\theta_2$ and $\Delta\omega$ librate around $0$ and the peak-to-valley amplitude of these librations are around $12$, $6$ and $6.1$ rad, respectively. \label{fig:2pfig100}}
\end{figure*}

In Section \ref{sec:2sec2}, we have discussed $r+1:r$ MMR case. 
We are now concentrating on the dynamics of 2:1 resonance of planets Kepler-62e and Kepler-62f orbiting Kepler 62. 
From Table \ref {tab:2p1}, it is clear that periods of Kepler-62e and Kepler-62f are 122.3874 and 267.291 d, respectively. So, there exists nearly 2:1 resonance between these two planets. In this case the two terms associated with the two first-order 2:1  arguments are
 $\theta_1=2\lambda_2-\lambda_1-\omega_1$ and $\theta_2=2\lambda_2
-\lambda_1-\omega_2$, where $\lambda_1$, $\lambda_2$ are the mean longitudes and $\omega_1, \omega_2$ are periastron longitudes of Kepler-62e and Kepler-62f, respectively. Also the apsidal lock between the orbiting companions is another important feature for the resonant systems. The relative apsidal longitudes is defined as $\Delta\omega=\omega_1-\omega_2$. If atleast one of the resonant angles among $\theta_1$ and $\theta_2$ librates around a constant value, then it is said to be in 2:1 MMR. Moreover, the system is said to be in apsidal co-rotation if $\Delta\omega$ also librates. In Fig. \ref{fig:2pfig100}, we depict the evolution of the resonant angles $\theta_1, \theta_2$ and $\Delta\omega$. From this figure, one can observe the behaviour of the resonant angles against time. We see that the two resonant angles $\theta_1$ and $\theta_2$ are librating about 0 rad, also $\Delta\omega$ librates around 0 rad. This results imply that the two planets Kepler-62e and Kepler-62f of Kepler-62 system are nearly 2:1 MMR and besides in apsidal co-rotation. It is also noticed that the peak-to-valley amplitude of libration of $\theta_1, \theta_2$ and $\Delta\omega$ are around 12, 6 and 6.1 rad, respectively. The resonance angle $\theta_1$ goes  to large amplitude oscillations. It may be due to the planet's large angular displacement from the periastron of its orbits \citep{ketchum2013mean}. 
\begin{figure*}
\includegraphics[height=0.35\textwidth]{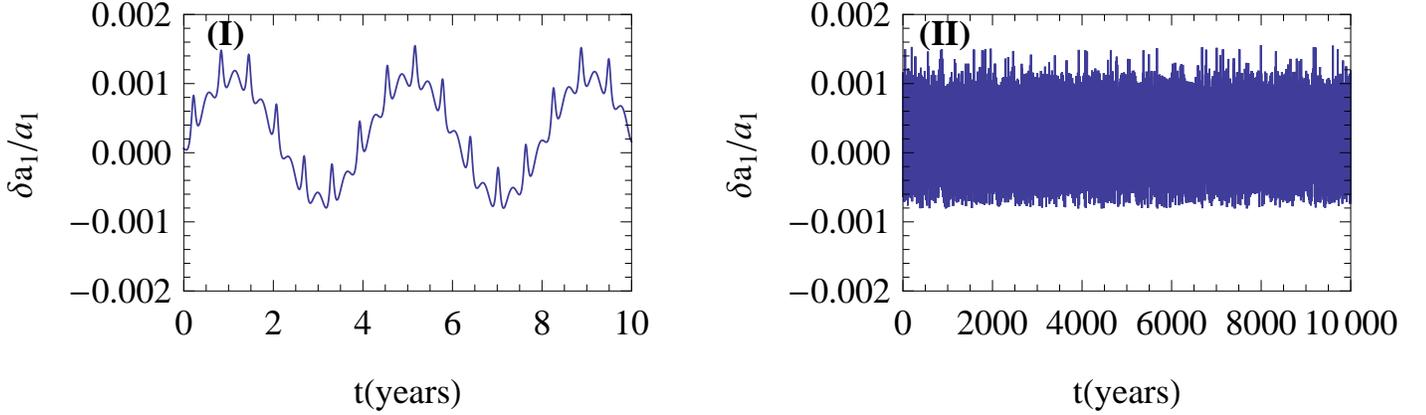}
\caption{Perturbative solution for the time variation of the semimajor axes: (I) for $t\in[0,10]$ , (II) for long time $t\in[0,10000]$ of inner planet Kepler-62e.\label{fig:2fig1}}
\end{figure*}
\begin{figure*}
\includegraphics[height=0.35\textwidth]{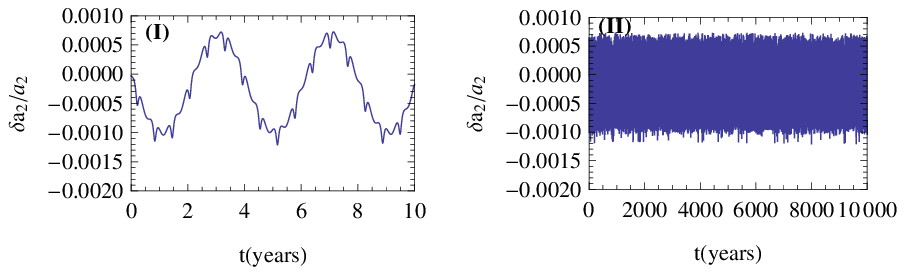}
\caption{Perturbative solution for the time variation of the semimajor axes: (I) for $t\in[0,10]$ , (II) for long time $t\in[0,10000]$ of outer planet Kepler-62f.\label{fig:2fig2}}
\end{figure*}
\begin{figure}
\includegraphics[height=0.6\textwidth]{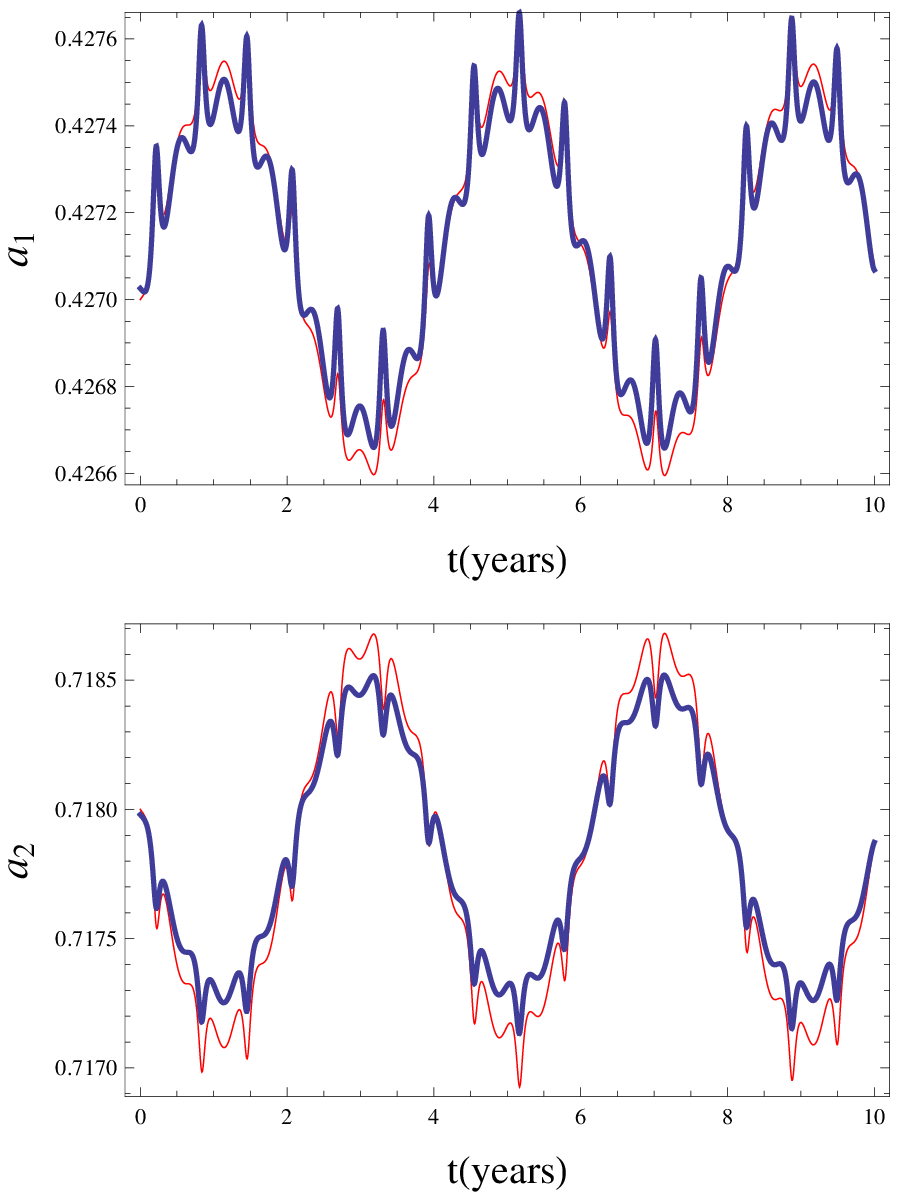}
\caption{Comparison between numerical and analytical solution for the time variation of the semimajor axes of Kepler-62 system: the thick line represents the result by analytical theory and the thin line represents the numerical solution.\label{fig:2fig3}}
\end{figure}
\begin{figure*}
\includegraphics[height=0.420\textwidth]{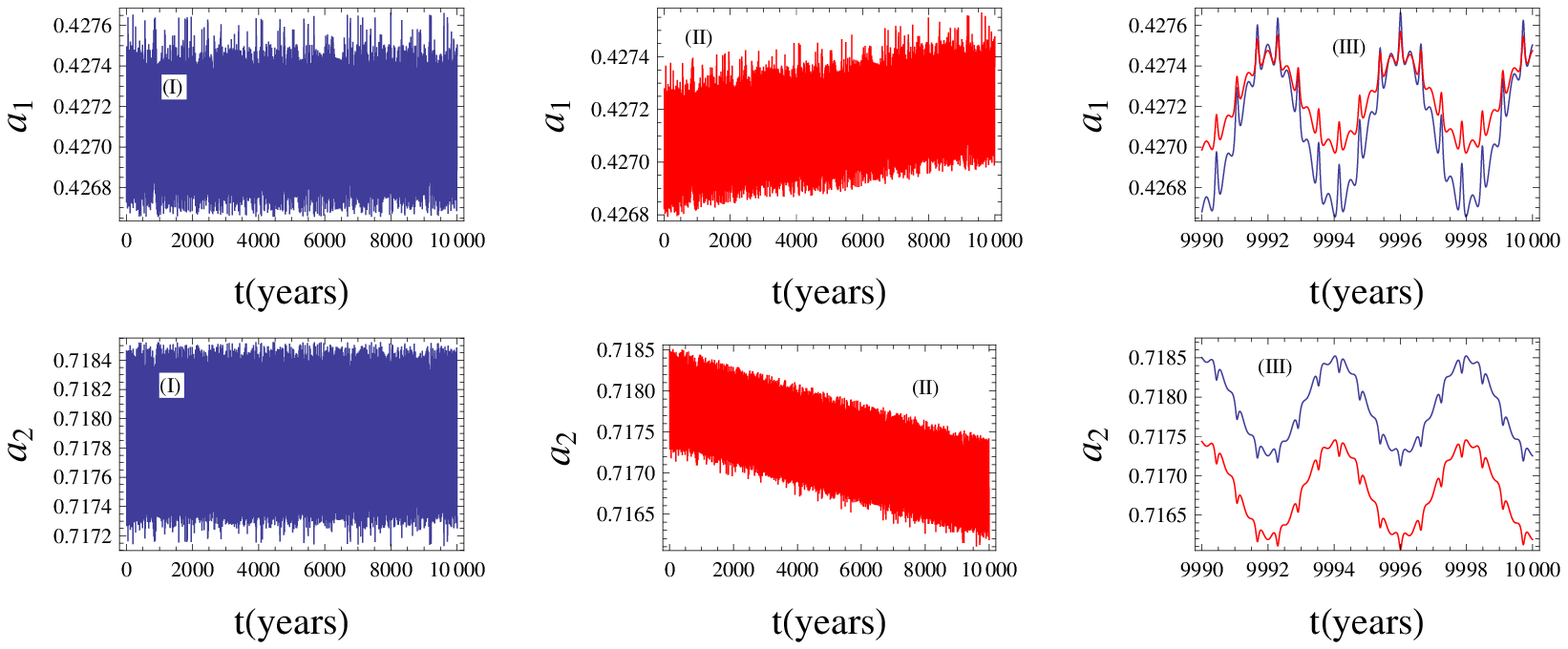}
\caption{Comparison between numerical and analytical solution of the semimajor axes of planets of Kepler-62 system for long time $t\in[0,10000]$. The upper panel corresponds to Kepler-62e and the lower panel corresponds to Kepler-62f. In each panel, (I) represents the result by analytical theory, (II) represents the numerical solution while (III) represents the zoom part of (I) 
and (II) for the time interval $t\in[9990,10000]$.\label{fig:2plongfig1}}
\end{figure*}
Now, if we put $r=1$ in Section \ref{sec:2sec2}, we can obtain the perturbation equations 
for the time variation of the orbital elements.
First we solve the equations for the semimajor axes. For this, we integrate equations \eqref{eq:2pa1} and \eqref{eq:2pa2} and then variations of the semimajor axes are given by $a_1(t)=a_{1,0}+\delta a_1(t)$,
\quad $a_2(t)=a_{2,0}+\delta a_2(t)$, where 
\begin{eqnarray}\label{eq:2p11}
\frac{\delta a_1(t)}{a_{1,0}}&=&\frac{2M_2 \alpha}{M_{\star}}\left(\frac{n_1}{n_1-n_2}\left[ Q(\psi(t),\alpha)-Q(\psi_0,\alpha)\nonumber \right.\right.\\ &&\left.\left.
-\alpha(\cos\psi(t)-\cos\psi_0)\right]-\frac{n_1}{2n_2-n_1}\left[-\left(2+\frac{\alpha}{2}D\right)
\right.\right.\nonumber\\&&\left.\left.
\times b_{\frac{1}{2}}^{(2)}(\alpha)e_1(\cos\theta_1(t)
-\cos\theta_{1,0})+\left(\frac{3}{2}+\frac{\alpha}{2}D\right)
\right.\right.\nonumber\\&&\left.\left.
\times b_{\frac{1}{2}}^{(1)}(\alpha)
e_2(\cos \theta_2(t)-\cos\theta_{2,0})\right]\right),
\end{eqnarray}
\begin{eqnarray}\label{eq:2eq24}
\frac{\delta a_2(t)}{a_{2,0}}&=&-\frac{2M_1 }{M_{\star}}\left(\frac{n_2}{n_1-n_2}\left[ Q(\psi(t),\alpha)-Q(\psi_0,\alpha)\nonumber \right.\right.\\ &&\left.\left.
-\alpha(\cos\psi(t)-\cos\psi_0)\right]-\frac{2n_2}{2n_2-n_1}\left[-\left(2+\frac{\alpha}{2}D\right)
\right.\right.\nonumber\\&&\left.\left.
\times b_{\frac{1}{2}}^{(2)}(\alpha)e_1(\cos\theta_1(t)
-\cos\theta_{1,0})+\left(\frac{3}{2}+\frac{\alpha}{2}D\right)b_{\frac{1}{2}}^{(1)}(\alpha)
\right.\right.\nonumber\\&&\left.\left.
\times e_2(\cos \theta_2(t)-\cos\theta_{2,0})\right]\right),
\end{eqnarray}
where
\begin{eqnarray}
&&\theta_j(t)=(2n_2-n_1)t+2(\sigma_2+\omega_2)-(\sigma_1+\omega_1)-\omega_j,\nonumber\\&&
\psi(t)=(n_1-n_2)t+(\sigma_1+\omega_1)-(\sigma_2+\omega_2).
\end{eqnarray}

Using the data given in Table \ref{tab:2p1}, we can determine $n_j=\frac{2\uppi}{P_j},\quad j=1,2,$ where $j=1$ means planet Kepler-62 and $j=2$ for Kepler-62f. Now from equations \eqref{eq:2p11} and \eqref{eq:2eq24}, we observe that there are two components for the variations in the semimajor axes. In the first component, period is equal to the period of the planets, $\frac{2\uppi}{|n_1-n_2|}\approx $225.757 d and corresponding fractional amplitude $\frac{M_jn_1}{M_{\star}|n_1-n_2|}\approx2\times10^{-4}$ and in the second component period is $\frac{2\uppi}{|2n_2-n_1|}\approx 1452.87$ days with fractional amplitude $\frac{M_jn_1}{M_{\star}|2n_2-n_1|}\approx1\times10^{-3}$. In Fig.\ref{fig:2fig1}, curve (I) represents the time variation of semimajor axis of planet Kepler-62e for the time interval $t\in{(0,10)}$ and curve (II) for long time $t\in(0,10000)$. For the same time interval, we have shown the time variation of the semimajor axis of planet Kepler-62f (in Fig.\ref{fig:2fig2}). Also, a comparison between numerical and analytical solution for the time variation of the semimajor axes is shown in Fig.\ref{fig:2fig3} to validate the result. The thick line represents result by analytical theory and the thin line represents numerical solution.
Also for long time $t\in(0,10000)$, a comparison between numerical and analytical solution for the time variation of the orbital semimajor axes of planets of Kepler-62 system are shown in Fig. \ref{fig:2plongfig1}. In this figure, the upper panel corresponds to Kepler-62e and while lower panel corresponds to Kepler-62f. In each panel, frame (I) and (II) represent the result by analytical theory and numerical solution while frame (III) shows the same behaviour of changes in semimajor axes for $t\in(9990,10000)$ (as in Fig. \ref{fig:2fig3}). We see that for long time, analytical solution of semimajor axis of Kepler-62e lies between (0.4267 and 0.4276) and numerical solution lies between (0.4268 and 0.4275) and in case of Kepler-62f analytical solution lies between (0.7172 and 0.7185) and numerical solution lies between (0.7164 and 0.7185).

\subsection{Secular resonance dynamics of Kepler-62 system}\label{sec:2}
Now we draw attention to the secular theory of Kepler-62 system by considering two planets, Kepler-62e and 
Kepler-62f, where the two planets are in $2:1$ MMR. We avoid the contributions from 
planets Kepler-62b, Kepler-62c and Kepler-62d because their contributions are much less than that of the 
mutual effect of two outer planets. Hence, we discuss secular resonance dynamics of planets Kepler-62e 
and Kepler-62f by ignoring the other planets.
In this case \begin{eqnarray}
&&E_1=-\frac{M_2}{M_\star}n_1\alpha\left(2+\frac{\alpha}{2}D\right)b_{\frac{1}{2}}^{(2)}(\alpha),\nonumber\\ &&
E_2=\frac{M_1}{M_\star}n_2\left(\frac{3}{2}+\frac{\alpha}{2}D\right)b_{\frac{1}{2}}^{(1)}(\alpha).
\end{eqnarray}
The solutions for the eccentricities can be written as
\begin{eqnarray}
&&p_j(t)=\sum_{i=1}^2e_{ji}\sin(g_it+\beta_i )+F_j\sin(2\lambda_2-\lambda_1),\nonumber\\&&
q_j(t)=\sum_{i=1}^2e_{ji}\cos(g_it+\beta_i )+F_j\cos(2\lambda_2-\lambda_1),
\end{eqnarray} 
where $B=[A-(2n_2-n_1)I]$. Using the theory as discussed in Section \ref{sec:2sec3}, we obtain two eigenfrequencies, $g_1=1.57246\times 10^{-3}$rad yr$^{-1} $and 
$g_2=2.65518\times 10^{-4}$rad yr$^{-1}$ together with $\beta_1=-1.27048$ rad, $\beta_2=1.14194$ rad, and $F_1=1.09302\times 10^{-3}, F_2=-1.29804\times 10^{-3}$, where $e_{ji}$ are given  in Table \ref{tab:11}.
\begin{figure}
\includegraphics[height=0.30\textwidth]{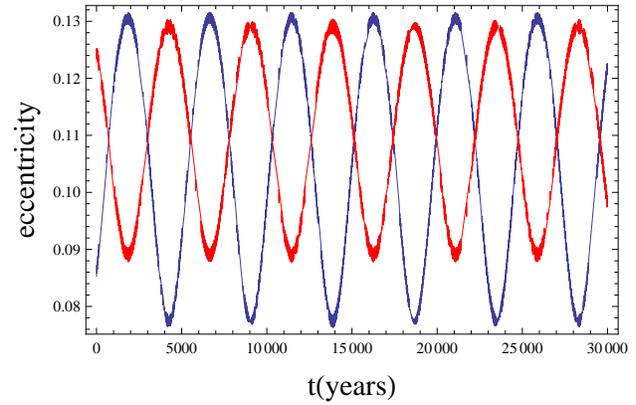}
\caption{Planet's eccentricity: curve (I) is eccentricity of Kepler-62e  , curve (II) is eccentricity 
of Kepler-62f for long time $t\in[0,30000]$.\label{fig:2pfig10}}
\end{figure}
The evolution of the eccentricities of the planets Kepler-62e and Kepler-62f are depicted in Fig. \ref{fig:2pfig10}. It is clear  that when one eccentricity reaches its maximum value, the other remains at its minimum value and conversely when one eccentricity is minimum, then other reaches exactly its minimum value. Also the eccentricity of Kepler-62e oscillates between $0.03230$ and $0.05391$ and eccentricity of Kepler-62f oscillates between $0.04154$ and $0.04787$.

\section{ HD 200964 system}
 \subsection{The 4:3 MMR}
\begin{figure*}
\includegraphics[height=0.21\textwidth]{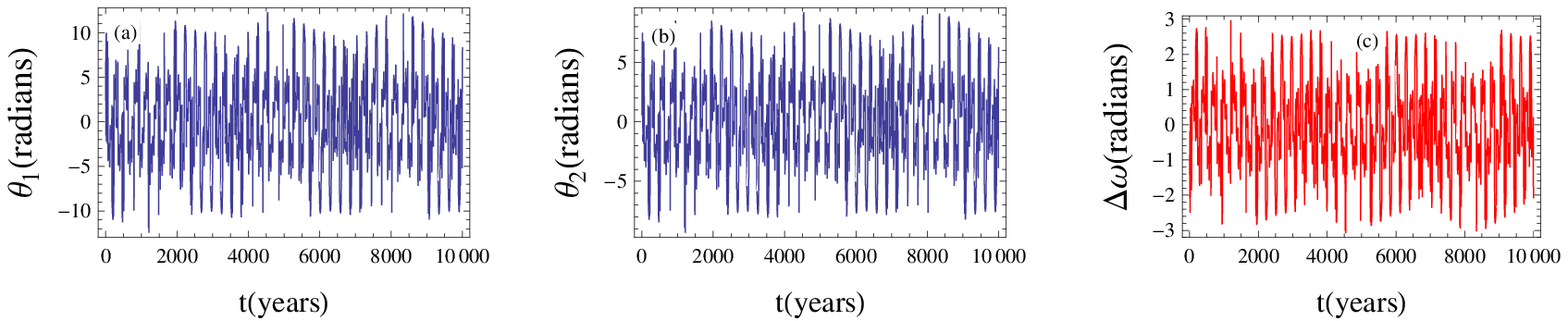}
\caption{The evolution of the resonant angles $\theta_1$ (a), $\theta_2$ (b) and apsidal angle $\Delta\omega$ (c) of HD 200964 system. Note that $\theta_1$, $\theta_2$ and $\Delta\omega$ librate around $0$ rad with an amplitude of $\pm11.34$, $\pm8$ and $\pm2.8$ rad, respectively. \label{fig:2pfig101}}
\end{figure*}

 For HD 200964 system data are taken from \citep{Johnson2011AJ....141...16J}. As in the previous 
 Section \ref{sec:2sec5}, we now focus on the dynamics of 4:3 resonance of planets HD 200964b and HD 200964c 
 orbiting the star HD 200964. In this case, also we may determine two terms associated with the two first-
 order 4:3 resonance with the arguments namely $\theta_3=4\lambda_2-3\lambda_1-\omega_1$ and $\theta_4=
 4\lambda_2-3\lambda_1-\omega_2$, and the relative apsidal longitude is $\Delta\omega=\omega_1-\omega_2$, where $\lambda_1$ and $\lambda_2$ are the mean longitudes of HD 200964b and HD 200964c, respectively. In Fig. \ref{fig:2pfig101}, curve (a) and curve (b)  show the behaviour of the resonant angles $\theta_1$ and $\theta_2$ and curve (c) represents the plot of the apsidal angle $\Delta\omega$ against time. From this figure, we see that $\theta_1$ librates around 0 rad with a larger amplitude of 
 $\pm11.34$ rad and $\theta_2$ librates around $0$ rad with an amplitude of $\pm8$ rad. This larger amplitude variations may be due to the planet's large angular displacements from the periastrons of its orbits \citep{ketchum2013mean}.  The libration of these two resonant angles confirm that there exists 4:3 MMR between the two planets of the system HD 200964. Also the apsidal angle $\Delta\omega$ librates about $0$ rad with an amplitude of $\pm2.8$ rad, which indicates that there exists an apsidal libration between two planets HD 200964b and HD 200964c.    
 
 \begin{figure*}
\includegraphics[height=0.32\textwidth]{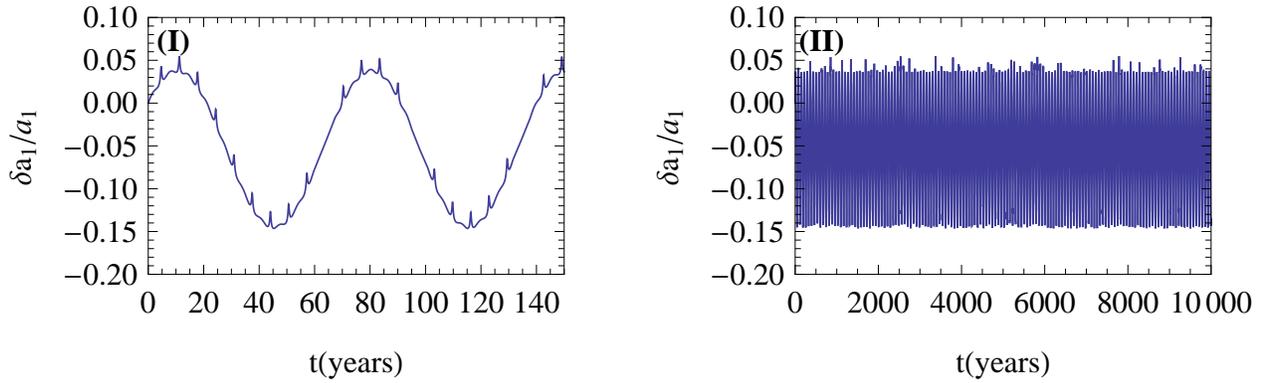}
\caption{Perturbative solution for the time variation of the semimajor axes: (I) for $t\in[0,150]$ , 
(II) for long time $t\in[0,10000]$ of inner planet HD 200964b.\label{fig:20fig1}}
\end{figure*}
\begin{figure*}
\includegraphics[height=0.35\textwidth]{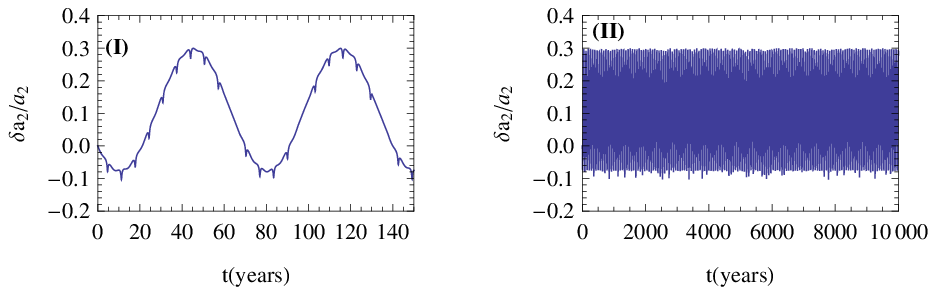}
\caption{Perturbative solution for the time variation of the semimajor axes: (I) for $t\in[0,150]$ , 
(II) for long time $t\in[0,10000]$ of outer planet HD 200964c.\label{fig:20fig2}}
\end{figure*}
\begin{figure}
\includegraphics[height=0.6\textwidth]{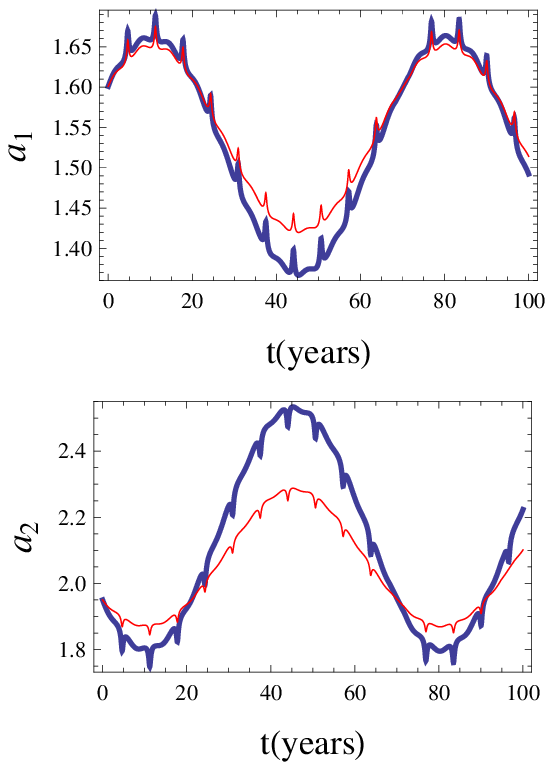}
\caption{Comparison between numerical and analytical solutions for the time variation of the semimajor axes of HD 200964 system: the thick line represents the result by analytical theory and the thin line represents the numerical solution.\label{fig:20fig3}}
\end{figure}
\begin{figure*}
\includegraphics[height=0.420\textwidth]{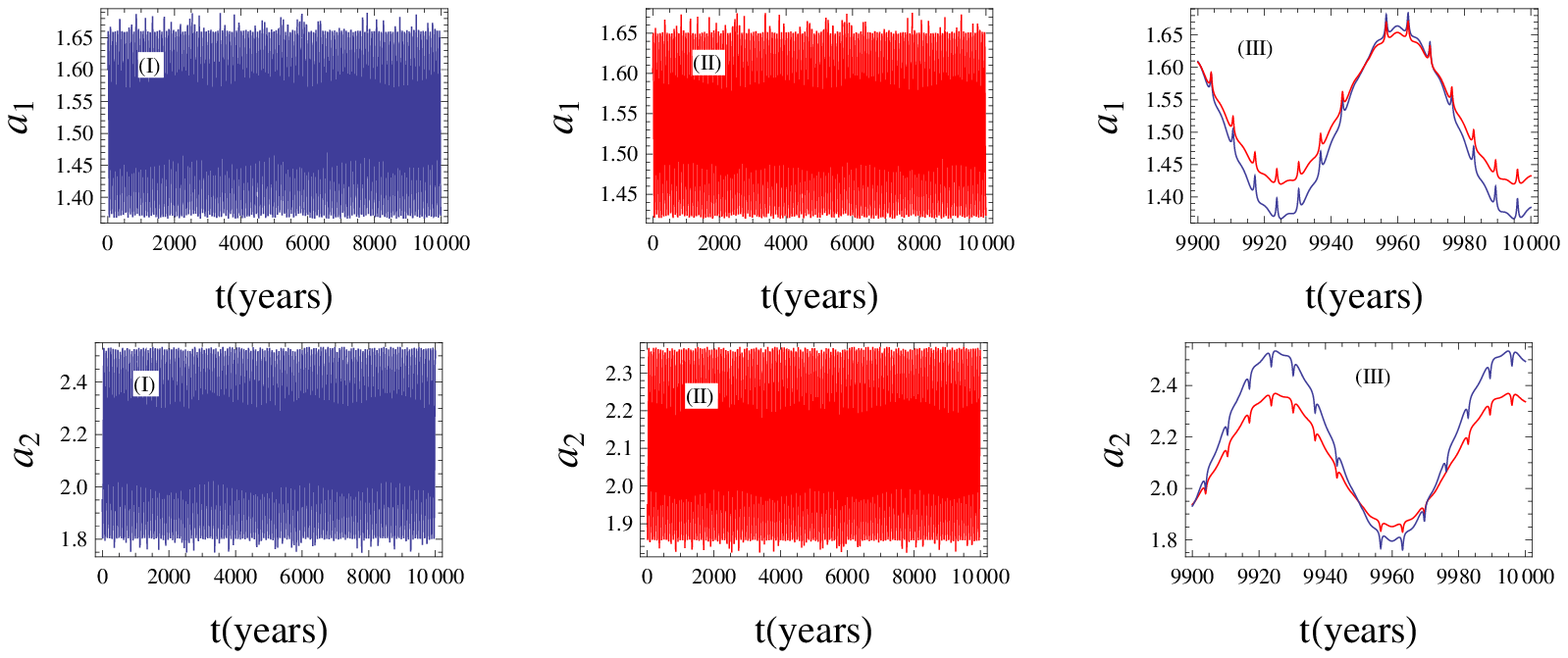}
\caption{Comparison between numerical and analytical solution of the semimajor axes of planets of HD 200964 system for long time $t\in[0,10000]$. The upper panel corresponds to HD 200964b and the lower panel corresponds to HD 200964c. In each panel, (I) represents the analytical solution, (II) represents the numerical solution, respectively, and (III) represents the zoom part of (I) 
and (II) for the time interval $t\in[9990,10000]$.\label{fig:2plongfig2}}
\end{figure*}
 
Now, the perturbation equations for the time variation of the semimajor axes can be obtained by substituting $r=3$ (in Section \ref{sec:2sec2}). 
We solve the equations for the semimajor axes then the variations of the semimajor axes are given by $a_1(t)=a_{1,0}+\delta a_1(t)$, \quad $a_2(t)=a_{2,0}+\delta a_2(t)$, where 
\begin{eqnarray}
\frac{\delta a_1(t)}{a_{1,0}}&=&\frac{2M_2 \alpha}{M_{\star}}\left(\frac{n_1}{n_1-n_2}\left[ Q(\psi(t),\alpha)-Q(\psi_0,\alpha)\nonumber \right.\right.\\ &&\left.\left.
-\alpha(\cos\psi(t)-\cos\psi_0)\right]-\frac{3n_1}{4n_2-3n_1}\left[-\left(4+\frac{\alpha}{2}D\right)
\right.\right.\nonumber\\&&\left.\left.\label{eq:2p29}
\times b_{\frac{1}{2}}^{(4)}(\alpha)e_1(\cos\theta_3(t)
-\cos\theta_{3,0})+\left(\frac{7}{2}+\frac{\alpha}{2}D\right)b_{\frac{1}{2}}^{(3)}(\alpha)
\right.\right.\nonumber\\&&\left.\left.
\times e_2(\cos \theta_4(t)-\cos\theta_{4,0})\right]\right),\\
\frac{\delta a_2(t)}{a_{2,0}}&=&-\frac{2M_1 }{M_{\star}}\left(\frac{n_2}{n_1-n_2}\left[ Q(\psi(t),\alpha)-Q(\psi_0,\alpha)\nonumber \right.\right.\\ &&\left.\left.
-\alpha(\cos\psi(t)-\cos\psi_0)\right]-\frac{4n_2}{4n_2-3n_1}\left[-\left(4+\frac{\alpha}{2}D\right)
\right.\right.\nonumber\\&&\left.\left.
\times b_{\frac{1}{2}}^{(4)}(\alpha)e_1(\cos\theta_3(t)
-\cos\theta_{3,0})+\left(\frac{7}{2}+\frac{\alpha}{2}D\right)b_{\frac{1}{2}}^{(3)}(\alpha)
\right.\right.\nonumber\\&&\left.\left.
\times e_2(\cos \theta_4(t)-\cos\theta_{4,0})\right]\right),
\end{eqnarray}
where
\begin{eqnarray}
&&\theta_j(t)=(4n_2-3n_1)t+4(\sigma_2+\omega_2)\nonumber\\
&&~~~~~~~~~~~~-3(\sigma_1+\omega_1)-\omega_j, \qquad (j=3,4)\\ 
&&\psi(t)=(n_1-n_2)t+(\sigma_1+\omega_1)-(\sigma_2+\omega_2).
\end{eqnarray}
\begin{table}
 \begin{minipage}{120mm}
\caption{Physical and orbital parameters of the HD $200964$ system \newline corresponding to the best
fit of \citep{Johnson2011AJ....141...16J}.\label{tab:2p2} }
  \begin{tabular}{@{}rrrrrrrrrrr@{}}
  \hline
  Parameter &HD $2600964$&HD $200964b$ & HD $200964c$&  \\\hline
$M_{\text{p}}\sin i$&1.44$M_{\odot}$& $1.85M_J $ &$0.895M_J$ \\ 
 Period(d)&&$613.8$&$825.0$\\           
 $a$(AU)&&$1.601$&$1.950$ \\ 
 $e$&&$0.040$&$0.181$\\
 $\omega$(deg)&&$288.0$&$182.6$\\
 $T_{\text{p}}$(JD)&&$2454900$&$2455000$\\
 \hline
\end{tabular}
\end{minipage}
\end{table}

We can determine $n_j=\frac{2\uppi}{P_j},\quad j=1,2$ using the data given in the Table \eqref{tab:2p2} where $j=1$ represents planet HD 200964b and $j=2$ for HD 200964c. For this case we observe that there are two components for the variations in the semimajor axes (see equation \eqref{eq:2p29}). In the first component, period is equal to the period of planets, $\frac{2\uppi}{|n_1-n_2|}\approx $2397.66 d and corresponding fractional amplitude $\frac{M_jn_1}{M_{\star}|n_1-n_2|}\approx2\times10^{-3}$. In  the second component, period is $\frac{2\uppi}{|4n_2-3n_1|}\approx 25575$ d with fractional amplitude $\frac{M_jn_1}{M_{\star}|2n_2-n_1|}\approx7\times10^{-2}$.  
In Fig.\ref{fig:20fig1}, curve (I) represents the time variation of semimajor axis of planet HD 200964b for the time interval $t\in{(0,150)}$ and curve (II) is for long time $t\in(0,10000)$. For the same time interval, we have shown the time variation of the semimajor axis of planet HD 200964c in Fig.\ref{fig:20fig2}. A comparison between numerical and analytical solution for the time interval $t\in(0,100)$ of the semimajor axes  are shown in Fig.\ref{fig:20fig3} to validate the result. The thick line represents results obtained by analytical theory and the thin line represents the numerical solution.
Also for long time $t\in(0,10000)$, a comparison between numerical and analytical solution for the time variation of the orbital semimajor axes of planets of HD 200964 system are shown in Fig. \ref{fig:2plongfig2}. In this figure, the upper panels are for HD 200964b and the lower panels are for HD 200964c. In each panel, frame (I) and (II) represent the result by analytical theory and numerical solution, respectively, while frame (III) shows the same behaviour of changes in semimajor axes for $t\in(9990,10000)$ as seen in Fig. \ref{fig:20fig3}. We see that for long time, analytical solutions of semimajor axis of HD 200964b lies between (1.37 and 1.66) and numerical solutions lies between (1.42 and 1.66) and while in the case of HD 200964c analytical solution lies between (1.8 and 2.5) and numerical lies between (1.84 and 2.4).

 \subsection{Secular resonance dynamics of HD 200964 system}
\begin{figure}
\includegraphics[height=0.30\textwidth]{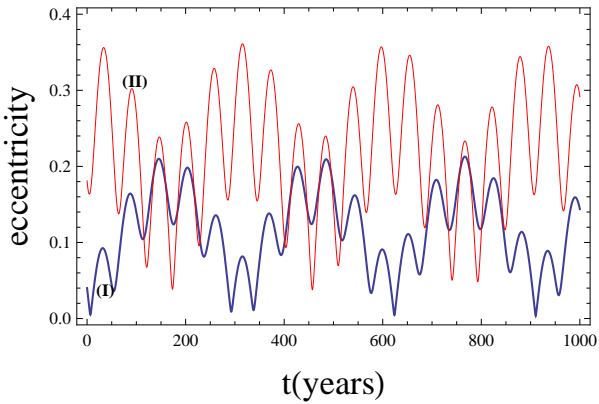}
\caption{Planet's eccentricity: curve (I) is eccentricity of HD 200964b  , curve (II) is eccentricity of
HD 200964c for long time $t\in[0,1000]$.\label{fig:fighd5}}
\end{figure}

As in Section \ref{sec:2}, we discuss the long-term variations of the eccentricities of the planets by
secular theory with MMR. Now we discuss the secular theory of HD 200964 system by 
considering all two planets. 
After some calculation, the solutions for the eccentricities can be written as 
\begin{eqnarray}
&&e_j\sin\omega_j=\sum_{i=1}^2e_{ji}\sin(g_it+\beta_i )+F_j\sin(4\lambda_2-3\lambda_1),\nonumber\\&&
e_j\cos\omega_j=\sum_{i=1}^2e_{ji}\cos(g_it+\beta_i )+F_j\cos(4\lambda_2-3\lambda_1),
\end{eqnarray} 
where the frequencies $g_i  \ (i=1,2)$ are the eigenvalues of the coefficient matrix $A$ and $e_{ji}$ 
are the components of the two corresponding eigenvectors. The normalization of the eigenvectors and the 
phases $\beta_i$ can be determined by the initial conditions. The amplitude of forcing will be same as 
in Eq.\eqref{eq:f} but B will be changed which is $B=[A-(4n_2-3n_1)I]$ and $E_1=-\frac{M_2}{M_{\star}}n_1\alpha\left(4+\frac{\alpha}{2}D\right)b_{\frac{1}{2}}^{(4)},\quad E_2=\frac{M_1}{M_{\star}}n_2\left(\frac{7}{2}+\frac{\alpha}{2}D\right)b_{\frac{1}{2}}^{(3)}.$
For this system, we obtain the eigenfrequencies, $g_1=2.16821\times10^{-2}$rad yr$^{-1}$ and $g_2=8.94711\times10^{-4}$rad yr$^{-1}$ together with $\beta_1=0.247833$, $\beta_2=0.502108$ and $F_1=4.3939\times10^{-2},\quad F_2=-9.84327\times10^{-2}.$
The evolution of the eccentricities of the two planets HD 200964b and HD 200964c are depicted in Fig. \ref{fig:fighd5}.
\begin{table}
 \begin{minipage}{120mm}
\caption{Values of $e_{ji}$ for the Kepler-62,
HD$200964$ and \newline ~~~~~~~~~~~~~~~~~~~~~~~~~~~~~~~Kepler-11 systems.}\label{tab:11} 
  \begin{tabular}{@{}rrrrrrrrrrr@{}}
  \hline
  System &$e_{ji}$&$i=1$ &$i= 2$&  \\\hline
Kepler-62&$e_{1i}$&0.026594  & 0.103945\\ 
 &$e_{2i}$&$-0.0200403$&0.109099\\           
 HD 200964&&& \\
&$e_{1i}$&$0.10331$&$-0.0655869$\\
 &$e_{2i}$&$-0.198811$&$-0.0638386$\\
  Kepler-11&&& \\
&$e_{1i}$&$0.00810542$&$0.043162$\\
 &$e_{2i}$&$-0.00229914$&$0.044661$\\
 \hline
\end{tabular}
\end{minipage}
\end{table}


\section{Kepler-11 system}
 \subsection{The 5:4 MMR}
 \begin{table}
 \begin{minipage}{120mm}
\caption{Physical and orbital parameters of the Kepler-11 system \newline corresponding to the best 
fit of \citep{lissauer2011closely}}. \label{tab:2p3}
  \begin{tabular}{@{}rrrrrrrrrrr@{}}
  \hline
  Parameter &Kepler-$11b$ & Kepler-$11c$&  \\\hline
Mass& $4.3$M$_{\oplus}$  &$13.5$M$_{\oplus}$ \\ 
 Period(d)&$10.30375$&$13.02502$\\           
 $a$(AU)&$0.091\pm0.003$&$0.106\pm0.004$ \\ 
 $e\cos\omega$&$0.0534\pm0.0383$&$0.0416\pm0.0332$\\
 $e\sin\omega$&$-0.0039\pm0.0072$&$-0.0007\pm0.0060$\\
 Epoch(BJD)&$2454971.5052$&$2454971.1748$\\
 \hline
\end{tabular}
\end{minipage}
\end{table}
 \begin{figure*}
\includegraphics[height=0.31\textwidth]{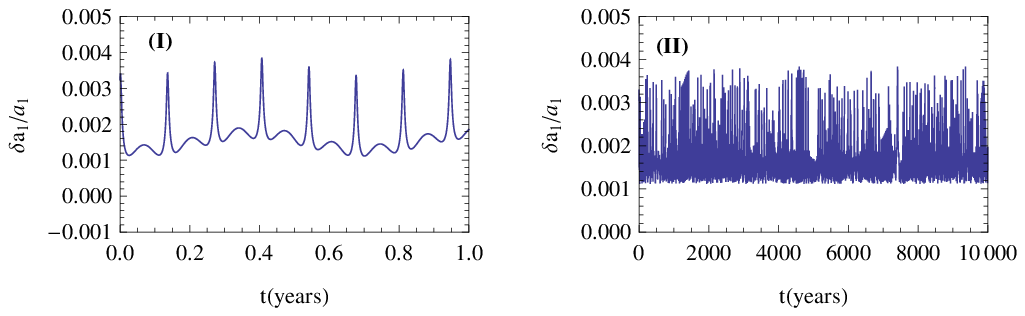}
\caption{Perturbative solution for the time variation of the semimajor axes: (I) for $t\in[0,1]$ , 
(II) for long time $t\in[0,10000]$ of inner planet Kepler-11b.\label{fig:30fig1}}
\end{figure*}
\begin{figure*}
\includegraphics[height=0.31\textwidth]{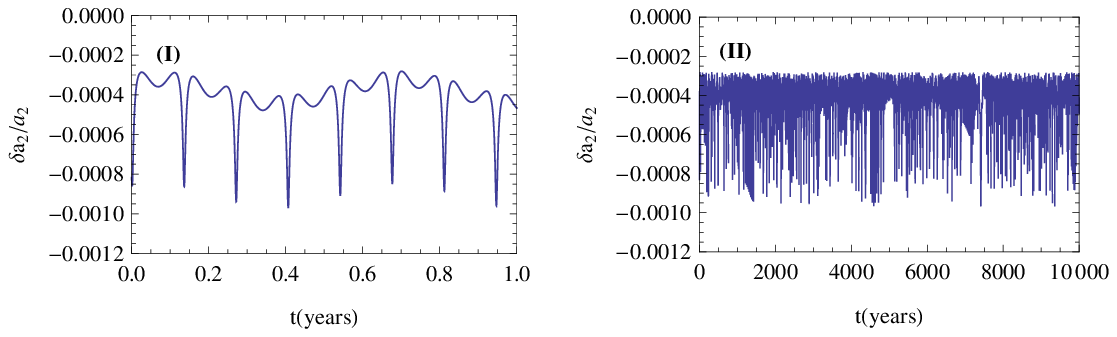}
\caption{Perturbative solution for the time variation of the semimajor axes: (I) for $t\in[0,1]$ , 
(II) for long time $t\in[0,10000]$ of outer planet Kepler-11c.\label{fig:30fig2}}
\end{figure*}

\begin{figure*}
\includegraphics[height=0.41\textwidth]{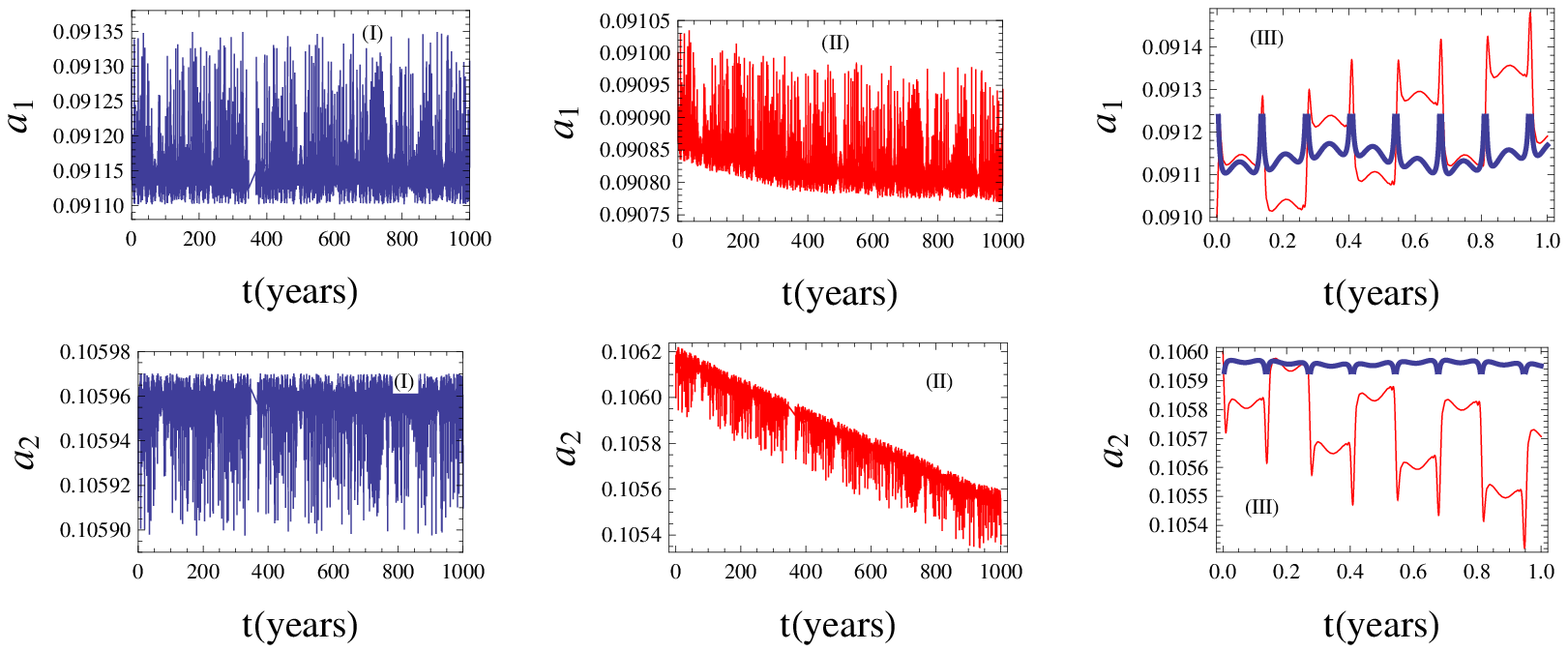}
\caption{Comparison between numerical and analytical solution of the semimajor axes of planets of Kepler-11 system for long time $t\in[0,10000]$. The upper panel corresponds to Kepler-11b and the lower panel corresponds to Kepler-11c. In each panel, (I) represents the result by analytical theory, (II) represents the numerical solution and (III) shows comparison between numerical and analytical solution for the time interval $t\in[0,1]$ of the semimajor axes of Kepler-11 system: the thick line represents the result by analytical theory and the thin line represents the numerical solution.\label{fig:2plongfig3}}
\end{figure*}
\begin{figure*}
\includegraphics[height=0.20\textwidth]{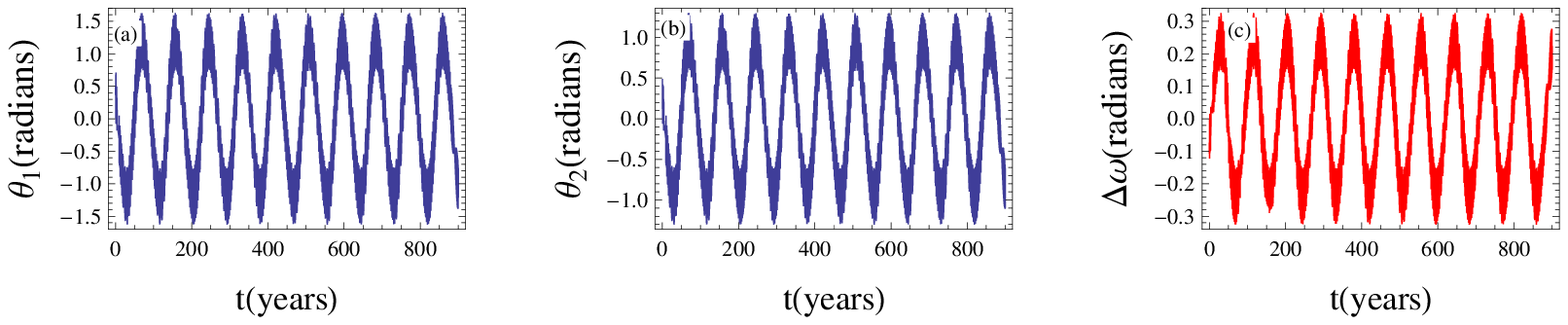}
\caption{The evolution of the resonant angles $\theta_1$ (a), $\theta_2$ (b) and apsidal angle $\Delta\omega$ (c) of Kepler-11 system. Note that $\theta_1$, $\theta_2$ and $\Delta\omega$ librate around $0$ rad with an amplitude of $\pm1.6$, $\pm1.3$ and $\pm0.32$ rad, respectively. \label{fig:2pfig103}}
\end{figure*}

 \cite{lissauer2011closely} observed perturbations of planets Kepler-11d and Kepler-11f by planet Kepler-11e  and confirm that all three sets of transits are produced by planets orbiting the same star and yield a somewhat weaker perturbations. The inner pair of observed planets, Kepler-11b and Kepler-11c, lie near a 5:4 orbital period resonance and strongly interact with one another. So, we avoid the contributions from planets Kepler-11d, Kepler-11e, Kepler-11f and Kepler-11g because their contribution are much less than that of the mutual effect of two inner planets.
We are now concentrating on the dynamics of 5:4 resonance of planets Kepler-11b and Kepler-11c orbiting
Kepler-11. In this case, the two terms associated with the two first-order 5:4  arguments are $\theta_1=5\lambda_2-4\lambda_1-\omega_1$ and $\theta_2=5\lambda_2
-4\lambda_1-\omega_2$, where $\lambda_1$ and $\lambda_2$ are the mean longitudes of Kepler-11b and
Kepler-11c, respectively. For this system, the plots of the resonant angles $\theta_1$, $\theta_2$ and the relative apsidal angle
against time are shown in curves (a), (b) and (c) (in Fig. \ref{fig:2pfig103}). Similarly, we see that for this system also $\theta_1$ librates around 0 rad with an amplitude of $\pm1.6$ rad and $\theta_2$ librates around 0 rad with an amplitude of $\pm1.3$ rad. Obviously, these results confirm that the two inner pair of observed planets, Kepler-11b and Kepler-11c of Kepler-1l system lie in 5:4 MMR. The libration of $\Delta\omega$ around 0 rad with an amplitude of $\pm0.32$ rad also means that the system is in apsidal co-rotation.    

Similarly, if we put $r=4$ in Section \ref{sec:2sec2}, we can obtain the perturbation equations 
for the time variation of the orbital elements.
 In this case, we also integrate Eqs.\eqref{eq:2pa1}
and \eqref{eq:2pa2} and then variations of the semimajor axes are given by $a_1(t)=a_{1,0}+\delta a_1(t)$,
\quad $a_2(t)=a_{2,0}+\delta a_2(t)$, where 
\begin{eqnarray}\label{eq:2p21}
\frac{\delta a_1(t)}{a_{1,0}}&=&\frac{2M_2 \alpha}{M_{\star}}\left(\frac{n_1}{n_1-n_2}\left[ Q(\psi(t),\alpha)-Q(\psi_0,\alpha)\nonumber \right.\right.\\ &&\left.\left.
-\alpha(\cos\psi(t)-\cos\psi_0)\right]-\frac{4n_1}{5n_2-4n_1}\left[-\left(5+\frac{\alpha}{2}D\right)
\right.\right.\nonumber\\&&\left.\left.
\times b_{\frac{1}{2}}^{(5)}(\alpha)e_1(\cos\theta_1(t)
-\cos\theta_{1,0})+\left(\frac{9}{2}+\frac{\alpha}{2}D\right)
\right.\right.\nonumber\\&&\left.\left.
\times b_{\frac{1}{2}}^{(4)}(\alpha) e_2(\cos \theta_2(t)-\cos\theta_{2,0})\right]\right),
\end{eqnarray}
\begin{eqnarray}
\frac{\delta a_2(t)}{a_{2,0}}&=&-\frac{2M_1 }{M_{\star}}\left(\frac{n_2}{n_1-n_2}\left[ Q(\psi(t),\alpha)-Q(\psi_0,\alpha)\nonumber \right.\right.\\ &&\left.\left.
-\alpha(\cos\psi(t)-\cos\psi_0)\right]-\frac{5n_2}{5n_2-4n_1}\left[-\left(5+\frac{\alpha}{2}D\right)
\right.\right.\nonumber\\&&\left.\left.
\times b_{\frac{1}{2}}^{(5)}(\alpha)e_1(\cos\theta_1(t)
-\cos\theta_{1,0})+\left(\frac{9}{2}+\frac{\alpha}{2}D\right)
\right.\right.\nonumber\\&&\left.\left.
\times b_{\frac{1}{2}}^{(4)}(\alpha)e_2(\cos \theta_2(t)-\cos\theta_{2,0})\right]\right),
\end{eqnarray}
where
\begin{eqnarray}
&&\theta_j(t)=(5n_2-4n_1)t+5(\sigma_2+\omega_2)-4(\sigma_1+\omega_1)-\omega_j,\nonumber\\&&
\psi(t)=(n_1-n_2)t+(\sigma_1+\omega_1)-(\sigma_2+\omega_2).
\end{eqnarray}
Using the numerical values of parameters given in Table \ref{tab:2p3}, we can calculate mean motion $n_j=\frac{2\uppi}{P_j},\quad j=1,2$ where $j=1$ means planet Kepler-11b and $j=2$ for Kepler-11c. Similar to the previous two systems in this case, we have also seen that there are two components for the variations in the semimajor axes. The period in the first component is equal to the period of the planets, $\frac{2\uppi}{|n_1-n_2|}\approx $49.3176 d and corresponding fractional amplitude $\frac{M_jn_1}{M_{\star}|n_1-n_2|}\approx2\times10^{-4}$ and in the second component period is $\frac{2\uppi}{|5n_2-4n_1|}\approx 230.861$ d with fractional amplitude $\frac{M_jn_1}{M_{\star}|2n_2-n_1|}\approx3\times10^{-3}$.

In Fig.\ref{fig:30fig1}, curve (I) represents the time variation of semimajor axis of planet Kepler-11b for the time interval $t\in{(0,1)}$ and curve (II) is for long time $t\in(0,10000)$. For the same time interval, we have shown the time variation of the semimajor axis of planet Kepler-11c (in Fig.\ref{fig:30fig2}). 
Comparison between numerical and analytical solution of the semimajor axes of planets of Kepler-11 system for long time $t\in(0,10000)$ are shown in Fig. \ref{fig:2plongfig3}. The upper panel is given for Kepler-11b and the lower panel is for Kepler-11c. In each panel, (I) represents the result by analytical theory, (II) represents the numerical solution and (III) shows comparison between numerical and analytical solution for the time interval $t\in[0,1]$ of the semimajor axes of Kepler-11 system. The thick line represents the result by analytical theory and the thin line represents the numerical solution.
We see that for long time, analytical solution of semimajor axis of Kepler-11b lies between (0.09110 and 0.09135) and numerical solution lies between (0.09079 and 0.09105) while in the case of Kepler-11c analytical solutions lies between (0.10590 and 0.10598) and numerical solution lies between (0.1054 and 0.1062).
Since we have used truncated disturbing function to first order in the eccentricities for the periodic terms and to second order for the secular terms neglecting the higher order terms, the discrepancies between the analytical and numerical results are expected (Fig.\ref{fig:2plongfig3}).
  
\subsection{Secular solution of Kepler-11 system}
\begin{figure}
\includegraphics[height=0.30\textwidth]{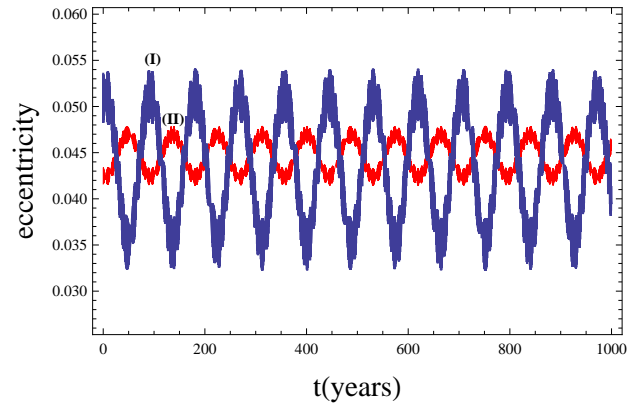}
\caption{Planet's eccentricity: curve (I) is eccentricity of Kepler-11b  , curve (II) is eccentricity of 
Kepler-11c for long time $t\in[0,500]$.\label{fig:2pfig5}}
\end{figure}

Now we discuss the secular theory of Kepler-11 system considering two planets Kepler-11b and 
Kepler-11c, where the two planets are in $5:4$ MMR.
In this case \begin{eqnarray}
&&E_1=-\frac{M_2}{M_\star}n_1\alpha\left(5+\frac{\alpha}{2}D\right)b_{\frac{1}{2}}^{(5)}(\alpha),\nonumber\\ &&
E_2=\frac{M_1}{M_\star}n_2\left(\frac{9}{2}+\frac{\alpha}{2}D\right)b_{\frac{1}{2}}^{(4)}(\alpha).
\end{eqnarray}
The solutions for the eccentricities can be written as
\begin{eqnarray}
&&p_j(t)=\sum_{i=1}^2e_{ji}\sin(g_it+\beta_i )+F_j\sin(5\lambda_2-4\lambda_1),\nonumber\\&&
q_j(t)=\sum_{i=1}^2e_{ji}\cos(g_it+\beta_i )+F_j\cos(5\lambda_2-4\lambda_1),
\end{eqnarray} 
also for this case $B=[A-(5n_2-4n_1)I]$.
With the numerical values (from Table \ref{tab:2p3}) and using the theory discussed in Section \ref{sec:2sec3}, for this system, we obtain $g_1=7.31037\times 10^{-2}$rad yr$^{-1} $and 
$g_2=1.53994\times 10^{-3}$rad yr$^{-1}$ together with $\beta_1=-0.397394$ rad, $\beta_2=-0.0312697$ rad, and $F_1=2.84641\times 10^{-3}, F_2=-9.39297\times 10^{-4}.$
The evolution of the eccentricities of the two planets Kepler-11b and Kepler-11c are depicted in Fig. \eqref{fig:2pfig5} over a time span of 1000 yr which is derived from the above solution for eccentricities. It is clear from the figure that the periodicity occurs in the variation of eccentricities. It is also shows that the minimum in eccentricity of Kepler-11b coincides with the maximum of Kepler-11c and reverse is also true. Moreover,  the eccentricity of Kepler-11b oscillates between $0.07633$ and $0.1313$ and eccentricity of Kepler-11c oscillates between $0.08769$ and $0.1302$.

 \section{Conclusions}
In this work we have used general three-body problem as a model for the study of dynamics of exoplanetary systems. We have applied the latest stability criteria given by \cite{Petrovich2015ApJ...808..120P} to examine the stability of exoplanetary systems Kepler-62, HD 200964, and Kepler-11. 
We have identified $(r+1):r$ MMR terms in the expression of disturbing function and obtained  the perturbations from the truncated disturbing function. It is found that the evolution of the resonant angles librates around constant value. Thus, according to this study, it is our opinion that 2:1, 4:3 and 5:4 near MMRs occur between Kepler-62e and Kepler-62f, HD 200964b and HD 200964c and Kepler-11b and Kepler-11c, respectively.
We have obtained the orbital solution of planets of Kepler-62 system in the 2:1 MMR, planets of HD 200964 in 4:3 MMR and planets of Kepler-11 in the 5:4 MMR. 
We have found that the periodicity occurs in the variation of eccentricities. It is also observed that a minimum in eccentricity of Kepler-62e coincides with a maximum of Kepler-62f and opposite is also true. Moreover maximum in Kepler-11b's eccentricity coincides with a minimum in Kepler-11c's eccentricity and conversely. A comparison is presented between the analytical solution and numerical solution.

Furthermore, the derived explicit analytical expressions would be used for the study of others newly discovered exoplanetary systems.
Moreover, the work would be extended  by considering additional planets (such as Kepler-62c, Kepler-11e) which negligibly perturb the pair of modelled planets in each system with the help of fully $N$-body simulations.
 
\section*{Acknowledgements}
We are thankful to Inter-University Centre for Astronomy and Astrophysics (IUCAA), Pune, India for supporting library visits
and for the use of computing facilities. Badam Singh Kushvah(BSK) is also grateful to the Indian
Space Research Organization (ISRO), Department of Space, Government of India, for providing financial support through the
RESPOND Programme (Project No -ISRO/RES/2/383/2012-13).

 \bibliographystyle{mn2e}

\label{lastpage}

\end{document}